%% file: main.tex
\documentclass[conference]{IEEEtran}
\IEEEoverridecommandlockouts
\usepackage{amsmath}
\usepackage{amssymb}
\usepackage{color,soul}
\usepackage{graphicx}
\allowdisplaybreaks[4]
\usepackage[noadjust]{cite}

\usepackage{float}
\usepackage{comment}
\usepackage{pgfplots}
\usepackage{xcolor}
\usepackage[font=small,skip=3pt]{caption}
  \pgfplotsset{compat=newest}
  \usetikzlibrary{plotmarks}
  \usetikzlibrary{arrows.meta}
  \usepgfplotslibrary{patchplots}
  \usepackage{grffile}
  \usepackage{url}
  \usepackage{enumerate}
  \usepackage{enumitem}
  \usepackage[flushleft]{threeparttable}
\usepackage[nolist,printonlyused]{acronym}
\usepackage{listings}
\newfloat{lstfloat}{t}{lop}
\floatname{lstfloat}{Code Block}

\usepackage{algorithm}
\usepackage{algorithmic}
\usepackage{url}
\usepackage{bm}
\usepgfplotslibrary{patchplots}
\renewcommand{\baselinestretch}{0.97}
\newcommand{\dc}{\!_\textup{\it DC}}
\usepackage{pgfkeys,pgfmath,pgfcore}
\pgfkeys{
   /textnumber/.style={
     /pgf/number format/.cd,
     fixed,use comma,fixed zerofill,precision=1,
     },
  }
\definecolor{mycolor4}{HTML}{A4978E}%
\definecolor{mycolor1}{HTML}{CD5C5C}
\definecolor{mycolor2}{rgb}{0.92941,0.69412,0.12549}%
\definecolor{mycolor3}{rgb}{0.23529,0.23529,0.23529}%
\begin{document}

\title{Grid-Coupled Dynamic Response of Battery-Driven Voltage Source Converters
\thanks{$^\ddagger$Corresponding author. This research was supported in part by the Director, Cybersecurity, Energy Security, and Emergency Response, Cybersecurity for Energy Delivery Systems program, of the U.S. Department of Energy, under contract DE-AC02-05CH11231.  Any opinions, findings, conclusions, or recommendations expressed in this material are those of the authors and do not necessarily reflect those of the sponsors of this work. This work was supported by the National Science Foundation under grants CPS-1646612 and CyberSEES-1539585. This work was also supported by the Laboratory Directed Research and Development (LDRD) at the National Renewable Energy Laboratory (NREL).}}%
\author{\IEEEauthorblockN{
Ciaran Roberts$^{\star,\ddagger}$, José Daniel Lara$^\dagger$, Rodrigo Henriquez-Auba$^{\star}$, \\ Bala Kameshwar Poolla$^\dagger$, and  Duncan S. Callaway$^{\star, \dagger}$%
}
\IEEEauthorblockA{$^\star$\textit{Department of Electrical Engineering and Computer Sciences}\\
$^\dagger$\textit{Energy and Resources Group} \\
\textit{University of California, Berkeley}\\
\textit{Berkeley, CA 94720} \\
\{ciaran\_r, jdlara, rhenriquez, bpoolla, dcal\}@berkeley.edu}
}%


\maketitle

\begin{abstract}
With the increasing interest in converter-fed islanded microgrids, particularly for resilience, it is becoming more critical to understand the dynamical behavior of these systems. This paper takes a holistic view of grid-forming converters and considers control approaches for both modeling and regulating the DC-link voltage when the DC-source is a battery energy storage system. We are specifically interested in understanding the performance of these controllers, subject to large load changes, for decreasing values of the DC-side capacitance. We consider a fourth, second, and zero-order model of the battery; and establish that the zero-order model captures the dynamics of interest for the timescales considered for disturbances examined. Additionally, we adapt a grid search for optimizing the controller parameters of the DC/DC controller and show how the inclusion of AC side measurements into the DC/DC controller can improve its dynamic performance. This improvement in performance offers the opportunity to reduce the DC-side capacitance given an admissible DC voltage transient deviation, thereby, potentially allowing for more reliable capacitor technology to be deployed.


\end{abstract}


\input{intro}

\input{models}

\input{method}

\input{results}
\input{conclusions}



\bibliographystyle{IEEEtran}
\bibliography{references}

\input{appendix}
%

\end{document}

%% file: intro.tex
\section{Introduction}
As synchronously connected power systems shift towards systems with high penetration of converter-interfaced generation (CIG), it becomes more critical to understand the dynamical and transient behavior of these systems. These converter-dominated power systems are already prevalent in the form of islanded microgrids, motivated by increased resilience to natural disasters \cite{chen2015resilient, li2017networked}. Recent work has explored the small-signal stability of the DC/AC converter and its interaction with the grid. A common approach when analyzing the voltage source converter (VSC) behavior is to model the DC-side of the converter as an ideal voltage source \cite{d2015small, UrosStability}. On the other hand, when studying the dynamics of the DC-side, the grid is often simplified as a resistive load \cite{bazargan2014stability, bazargan2018reduced}. From a small-signal perspective, an independent analysis of each subsystem separately may be adequate due to the minimal interaction of their control loops. However, this approach gives little insight into the dynamical behavior of these coupled systems during grid-scale transient events, particularly, faults or large load steps and when the operating conditions differ substantially from the steady-state operating point used in the linearization. 

This paper explores the performance of the DC-link capacitor of a battery energy storage system (BESS) subject to AC-side disturbances, under different DC-side control strategies. The objective of these control loops on the DC-side is to tightly regulate the DC voltage across a DC-link capacitor. The DC-link capacitors act as energy buffers and support a constant voltage on the DC-side of the CIG. A tight regulation of this voltage is critical to the operation of the CIG, as momentary drops in this voltage restrict the VSC's power production capabilities \cite{bazargan2018reduced}. Therefore, large electrolytic capacitors are used in order to have a substantial buffer to minimize the DC voltage deviations during disturbances. These capacitors being typically bulky, expensive, unreliable are one of the most common modes of failure in power electronic systems \cite{wang2013toward}- with system transients and overloading identified as two of the primary causes of failure \cite{yang2011industry}. 

One proposed improvement in converter design is to replace these electrolytic capacitors with small film capacitors that are more robust and reliable \cite{lee2013dc}. As the DC-link capacitance is reduced, voltage fluctuations during transients increase as there is a momentary mismatch between the power injected into the grid and the power supplied from the DC source e.g., a battery. In order to deploy these small film capacitors, the DC-side control must rapidly correct any difference between these currents to ensure adequate AC-side operation and minimize transient over-voltages on the capacitor.

In this work, we examine different control approaches for minimizing the required DC-link capacitance of a BESS. Specifically, we consider the case of a grid-forming inverter supporting an islanded microgrid with a BESS as its DC source. Grid-forming inverters differ from grid-following inverters--the dominant mode of operation today, in that the former behave as a controllable voltage source behind a coupling reactance \cite{elkhatib2018evaluation}. Consequently, they do not directly control their power injection into the grid but rather control the frequency and amplitude of their output voltage \cite{UrosStability}. Their power injections, therefore, inherently increase or decrease to balance any changes in load. When choosing a DC-link capacitor to regulate the DC voltage of a grid-forming inverter, adequate care must be taken that it is appropriately sized to ensure satisfactory behavior under the largest expected load change and/or fault conditions, thus presenting challenges in the sizing of the DC-link capacitor for grid-forming converters. This work considers the existing measurements used in the control loop of grid-forming converters as inputs into the DC/DC controller to predict the evolution of DC-link capacitance dynamics and consequently, improve the regulation of the DC bus voltage.

For modeling our DC source, we consider a Li-ion battery as the BESS. In comparison to previous work which modeled the battery as an ideal voltage behind a resistor \cite{bazargan2014stability, bazargan2018reduced}, we employ a model of the battery which captures the dynamics of the electrochemical processes as we increase/decrease the current drawn from the battery. Furthermore, as we reduce the DC-link capacitance and the dynamics on the DC-side become faster, it may become more important to model the underlying battery dynamics to accurately capture the dynamical response of the DC source \cite{ferraz2018high}. 

The contributions of this paper are as follows: 
\begin{enumerate}[leftmargin=0.1cm, itemindent=0.5cm]
\setlength{\itemsep}{1.5pt}
    \item we develop a full-order dynamical model for a battery-driven voltage source converter,
    \item we examine the impact of battery chemistry dynamics on overall DC-side dynamical response and establish that a zero-order model captures the dynamics of interest for the disturbances considered,
    \item we improve upon the DC-side controller in \cite{bazargan2018reduced} by the inclusion of AC-side measured quantities to predict evolution of DC-side dynamics to compensate for the DC/DC controller dead-time and DC/DC inductor dynamics,
    \item we show that, for particular parameterizations of inner-control loops, the behavior of the VSC can help reduce the risk of saturation of the VSC modulation index.  
\end{enumerate}


%% file: models.tex
\section{System Modeling and Control Implementation} \label{sec:model}

\subsection{Grid Forming VSC Control Scheme}
\label{subsec:VSC}

\begin{figure}[h]
\includegraphics[width=0.47\textwidth]{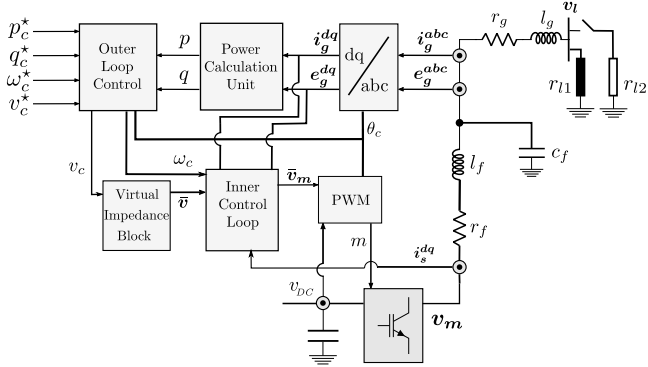}
\caption{Grid-forming VSC control scheme.}
\label{fig:vsc}
\end{figure}

The modeling and simulation of the AC-side, including the VSC, is implemented in a Synchronous Reference Frame (SRF), with the mathematical model defined in per unit. The $(\textup{\it dq})$-frame quantities are represented in bold, lower-case complex space vectors of the form: $\boldsymbol{x} = x^d+jx^q$. The proposed control model depicted in Fig.~\ref{fig:vsc} is based on a state-of-the-art VSC control scheme described in \cite{DArco2014,UrosISGTeurope,UrosEnergyCon}. The power calculation unit computes the active and reactive quantities given by $p_c+jq_c=\boldsymbol{e}_g\bar{\boldsymbol{i}_g}$ where $(\bar{.})$ denotes the complex conjugate. This is followed by an outer control loop that consists of active and reactive power controllers providing the output voltage magnitude $v_c$ and frequency $\omega_c$ references by adjusting the predefined set points $(x^{\star})$ according to a measured power imbalance:
\begin{equation}\label{eq:outer_loop}
    \omega_{c} = \omega_{c}^{\star} + R_{c}^p\, (p_{c}^{\star}-\tilde{p}_c),\quad
    v_{c} = v_{c}^{\star} + R_{c}^q\, (q_{c}^{\star}-\tilde{q}_c),
\end{equation}
where $R_{c}^p$, $R_{c}^q$ denote the active, reactive power droop gains and $\tilde{p}_c$, $\tilde{q}_c$ represent the low-pass filtered active, reactive power measurements of the form:
\begin{equation}
        \dot{\tilde{p}}_c = \omega_z\, (p_c-\tilde{p}_c), \quad \dot{\tilde{q}}_c = \omega_z\, (q_c-\tilde{q}_c),
\end{equation}
where $\omega_z$ is the filtering frequency. The outer-loop voltage set point may be passed through a virtual impedance block $(r_v,l_v)$, resulting in a  cross-coupling between the $d$- and $q$-components via a terminal current measurement $\boldsymbol{i}_g$ as
\begin{equation}\label{eq:virtual_impedance}
    \bar{\boldsymbol{v}} = v_c - (r_v + j\omega_c\, l_v)\, \boldsymbol{i}_g.
\end{equation}
This new voltage vector set point and the frequency set point are then fed to the inner control loop consisting of cascaded voltage and current controllers operating in a SRF
\begin{subequations}\label{eq:vsc_inner_loops}
\begin{align}
    \boldsymbol{\bar{i}}_s&=K_p^v\,(\bar{\boldsymbol{v}}-\boldsymbol{e}_g)+K_i^v\,\boldsymbol{\xi}+j \omega_c\,c_f\, \boldsymbol{e}_g+K_f^i\,\boldsymbol{i}_g, \label{eq:srf_v} \\
    \bar{\boldsymbol{v}}_m&=K_p^i\,(\bar{\boldsymbol{i}}_s-\boldsymbol{i}_s)+K_i^i\,\boldsymbol{\gamma}+j\omega_{c}\, l_f\, \boldsymbol{i}_s+K_f^v\,\boldsymbol{e}_g, \label{eq:srf_i}
\end{align}
\end{subequations}
where $\dot{\boldsymbol{\xi}}=\bar{\boldsymbol{v}}-\boldsymbol{e}_g$ and $\dot{\boldsymbol{\gamma}}=\bar{\boldsymbol{i}}_s-\boldsymbol{i}_s$ denote the respective integrator states; $\bar{\boldsymbol{i}}_s$ and $\bar{\boldsymbol{v}}_m$ represent the internally computed current and voltage references, $\boldsymbol{e}_g$ is the voltage measurement at the converter terminal to the grid, $\boldsymbol{i}_s$ is the switching current, $K_p$, $K_i$, and $K_f$ are the proportional, integral, and feed-forward gains respectively, and superscripts $v$ and $i$ denote the voltage and current SRF controllers. The output voltage reference $\bar{\boldsymbol{v}}_m$ combined with the DC-side voltage $v_{\dc}$ generates the Pulse-Width Modulation (PWM) signal $\boldsymbol{m}$. 

The electrical interface to the microgrid includes an $\textup{\it RLC}$ filter $(r_f, l_f, c_f)$ and an equivalent impedance $(r_g, l_g)$ modeled in SRF and defined by the angular converter frequency
\begin{subequations}
\begin{align}
    \dot{\boldsymbol{i}}_s&=\frac{\omega_{b}}{l_f}(\boldsymbol{v}_m-\boldsymbol{e}_g)-\left(\frac{r_f}{l_f}\,\omega_{b}+j\omega_{b} \,\omega_c\right)\boldsymbol{i}_s, \label{eq:elSys1}\\
    \dot{\boldsymbol{i}}_g&=\frac{\omega_b}{l_g}(\boldsymbol{e}_g-\boldsymbol{v}_l)-\left(\frac{r_g}{l_g}\,\omega_b+j\omega_b\, \omega_c\right)\boldsymbol{i}_g, \label{eq:elSys2}\\
    \dot{\boldsymbol{e}}_g&=\frac{\omega_{b}}{c_f}(\boldsymbol{i}_s-\boldsymbol{i}_g)-j\omega_c\, \omega_b \,\boldsymbol{e}_g, \label{eq:elSys3}
\end{align}
\end{subequations}
with $\boldsymbol{v}_m$ representing the modulation voltage and $\boldsymbol{v}_l$ denoting the nodal voltage at the load bus. The system base frequency is represented by $\omega_b$ and equals the nominal frequency. The complete state-space representation of a single grid-forming inverter, therefore, comprises 13 states of the form
\begin{align}
    \hat{x}_\textup{\it vsc}&=\left[\boldsymbol{e}_g^\textup{\it dq},\,\boldsymbol{i}_g^\textup{\it dq},\,\boldsymbol{i}_s^\textup{\it dq},\,\boldsymbol{\xi}^\textup{\it dq},\,\boldsymbol{\gamma}^\textup{\it dq},\,\theta_c,\,\tilde{p}_c,\,\tilde{q}_c\right]^\top.
\end{align}

The control input vector $u_\textup{\it vsc}=\left[p_c^{\star},\,q_c^{\star},\,v_c^{\star},\omega_c^{\star}\right]^\top$ provides operator set points. More details on the overall converter control structure and employed parametrization can be found in \cite{UrosEnergyCon,UrosISGTeurope,UrosStability}.

\subsection{DC-side model} \label{subsec:dc_side}

\begin{figure}[h]
\centering
\includegraphics[width=0.47\textwidth]{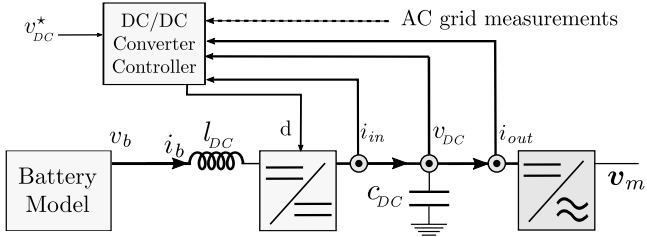}
\caption{DC-side model.} 
\label{fig:dc_side_model}
\end{figure}

The modeling of the DC-side consists of a BESS, an idealized DC/DC buck/boost converter with an appropriately sized inductor, and a DC-link capacitor. This interconnected system is then interfaced to the VSC as shown in Fig. \ref{fig:dc_side_model}.

\subsubsection{DC/DC Controller}
\label{subsec:dc_control}

\begin{figure}[h]
\includegraphics[width=0.47\textwidth]{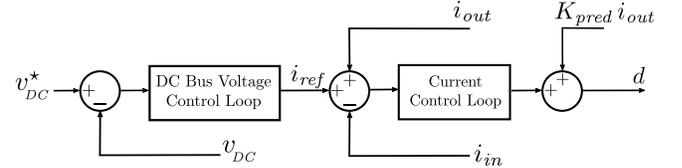}
\caption{Structure of the DC-side controller}
\label{fig:dc_side_control}
\end{figure}

For the DC/DC controller in Fig.~\ref{fig:dc_side_model}, we investigate the improved dynamical performance with the inclusion of the measured AC-side quantities into the control logic. A dual-loop PI  DC/DC controller is shown in Fig. \ref{fig:dc_side_control} and modeled as
\begin{subequations}\label{eq:dc_dc_controller}
\begin{align}
        &\dot{\eta} = v_{\dc}^{\star}-v_{\dc} \label{eq:dc_outer_a} \\ 
        &i_\textup{\it ref} = K_{p}^{v_{\dc}}\,(v_{\dc}^{\star}-v_{\dc}) +  K_{i}^{v_{\dc}}\,\eta \label{eq:dc_iref},\\
        &\dot{\zeta} = i_\textup{\it ref}+i_\textup{\it out}-i_\textup{\it in}, \label{eq:iref_int} \\ 
        &d = K_{p}^{i_{\dc}}\,(i_\textup{\it ref}+i_\textup{\it out}-i_\textup{\it in}) +  K_{i}^{i_{\dc}}\,\zeta + K_\textup{\it pred}\,\Delta i_\textup{\it out}. \label{eq:dc_d} 
\end{align}
\end{subequations}

The outer-loop \eqref{eq:dc_outer_a}-\eqref{eq:dc_iref}, maintains a constant DC bus voltage while the inner loop \eqref{eq:iref_int}-\eqref{eq:dc_d}, is for current tracking. The inclusion of a feed-forward term $i_\textup{\it out}$, in the internal PI control loop is for improving the controller performance by the addition of information about the disturbance. This disturbance was primarily a set-point change of the VSC in previous works \cite{bazargan2014stability}. For the case of a grid-forming VSC, however, this disturbance includes unexpected load changes where the additional required power will be inherently drawn from the DC-link capacitor. 

The addition of the term $K_\textup{\it pred}\,\Delta i_\textup{\it out}$ in \eqref{eq:dc_d} is motivated by \cite{gu2006dc}, where the authors sought to minimize the required DC-link capacitance for a converter-interfaced three-phase load. In \cite{gu2006dc} the authors note that the inclusion of a feed-forward term alone is inadequate to instantaneously balance the current flow across the capacitor due to inherent system response time delays, mainly due to inductor dynamics. To offset these delays, we use a one-step predictor based on the forward Euler method to predict the evolution of system dynamics. The feed-forward predicted current, $\Delta i_\textup{\it out}$ value is approximated by \eqref{eq:predict_current}
\begin{equation} \label{eq:predict_current}
    \Delta i_\textup{\it out} \approx \dfrac{\Delta P}{\Delta v_{\dc}}\approx \dfrac{T_{s}(v_{m}^{d}\,\dot{i_{s}^{d}} + v_{m}^{q}\,\dot{i_{s}^{q}})}{v_{\dc}},
\end{equation}
where $T_{s}$ is the switching period of the DC/DC converter, $\dot{i_{s}^{d}}$ and $\dot{i_{s}^{q}}$ are calculated using \eqref{eq:elSys1}. We benchmark the improvement in dynamical performance for a non-zero $K_\textup{\it pred}$ against the controller in \cite{bazargan2014stability}. The advantage of a one-step predictor over derivative control in a PID controller is that we can predict the evolution of the DC-side dynamics before they begin to manifest and minimize noise amplification in estimating the rate of change of the current. The duty-cycle $d$ of the DC/DC converter in this work has a maximum value of $0.9$ to mimic the behavior of a practical converter \cite{zhao2006dc}. 

\subsubsection{Battery Model} 
\begin{figure}[h]
\includegraphics[width=0.47\textwidth]{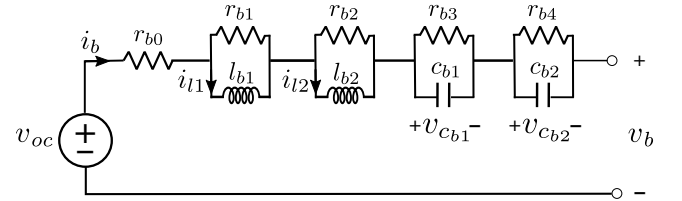}
\caption{A generalized $4^\textup{\it th}$-order battery model.} 
\label{fig:general_battery}
\end{figure}

As previously outlined, prior work on this topic modeled the electrochemical battery as an ideal voltage source behind a resistor \cite{bazargan2014stability, bazargan2018reduced}. In the presence of a large DC-link capacitance and consequently a large energy buffer, this is a reasonable modeling assumption. However, as we reduce the DC-link capacitance, the dynamics of the electrochemical storage may become more important to model. A common method for parameterizing an equivalent circuit model for batteries is electrochemical impedance spectroscopy \cite{ferraz2018high, uno2011influence}. This method measures the voltage response to harmonic current input across a frequency range of interest ($3$ kHz to $30$ kHz \cite{rahmoun2016mathematical}) and an equivalent circuit is adapted to this data. These experimental data show that at high frequencies ($\geq 250-400$ Hz) the battery exhibits inductive behavior while lower frequencies ($\leq 250-400$ Hz) have a more capacitive response \cite{waag2013experimental, ferraz2018high, rahmoun2016mathematical}. A generalized battery is shown in Fig.~\ref{fig:general_battery} where the high frequency behavior is modeled by a series of 2 $\textup{\it RL}$ parallel branches and the low frequency behavior is modeled by a series of 2 $\textup{\it RC}$ parallel branches. 

Within this work we combine the two-time constant $RC$ battery model from \cite{hart2014modeling} with the two-time constant $RL$ model from \cite{rahmoun2016mathematical} as shown in Fig.~\ref{fig:dc_side_model}. Both of these batteries' chemistries are based on Lithium-ion and offer reasonable initial parameterization of a dynamic BESS model.
\subsubsection{DC-side Electrical Model}
In practice, the DC/DC converter is a buck/boost converter capable of both charging and discharging the battery. Here, we focus on the case when the converter is operating in the boost mode, i.e., supplying power to the grid. A similar analysis holds for the buck mode of operation. The per-unit averaged equations governing the electrical behavior on the DC-side with the converter operating in continuous mode, similar to \cite{wang2014nonlinear}, are then given by
\begin{subequations}\label{eq:dc_elec_eqs}
\begin{align}
    \dot{i}_{l1} &=\dfrac{\omega_{b}}{l_{b1}}(r_{b1}(i_{b}-i_{l1})), \quad
    \dot{i}_{l2} = \dfrac{\omega_{b}}{l_{b2}}(r_{b2}(i_{b}-i_{l2})), \\
    \dot{v}_{c_{b1}} &= \dfrac{\omega_{b}}{c_{b1}}\left(i_{b}-\dfrac{v_{c_{b1}}}{r_{b3}}\right),\quad
    \dot{v}_{c_{b2}} = \dfrac{\omega_{b}}{c_{b2}}\left(i_{b}-\dfrac{v_{c_{b2}}}{r_{b4}}\right),\\
    v_{b} &= v_{oc}-i_{b}\,r_{b0}-r_{b1}(i_{b}-i_{l1})-r_{b2}(i_{b}-i_{l2}) \\
    &- v_{c_{b1}} - v_{c_{b2}} \\
    \dot{i}_{b} &= \dfrac{\omega_{b}}{l_{\dc}}(v_{b} - (1-d)\,v_{\dc}),\\
    \dot{v}_{\dc} &= \dfrac{\omega_{b}}{c_{\dc}}(i_\textup{\it in}-i_\textup{\it out}),\\
    v_{\dc}i_\textup{\it in}&=v_{b}\,i_{b},\label{eq:dc_current}
    \end{align}
\end{subequations}
where $\omega_{b}$ is the AC base frequency, $d$ is the duty-cycle of the DC/DC converter, further discussed in Section~\ref{subsec:dc_control}, and $i_\textup{\it out}$ is the current flowing into the AC grid and given by
\begin{equation}
    i_\textup{\it out} = \frac{p_\textup{\it inv}}{v_{\dc}} = \dfrac{v_{m}^{d}\,i_{s}^{d} + v_{m}^{q}\,i_{s}^{q}}{v_{\dc}}.
\end{equation}

The full state-space model of the DC-side with a $4^\textup{\it th}$-order dynamic BESS model, denoted by  $\hat{x}^{4^\textup{\it th}}_{\dc}$, is given by
\begin{equation}\label{eq:dc_full_states}
    \hat{x}^{4^\textup{\it th}}_{\dc}=\left[i_{l1},\,i_{l2},\,v_{c_{b1}},\,v_{c_{b2}},\,i_{b},\,v_{\dc},\,\eta,\,\zeta \right]^\top,
\end{equation}
with the control input $u_{\dc}=v_{\dc}^{\star}$. The $2^\textup{\it nd}$-order model of the DC-side neglects the inductor dynamics of the battery (i.e., retains only the 2 $\textup{\it RC}$ branches in Fig.~\ref{fig:general_battery}), while the $0^\textup{\it th}$-order model further neglects the dynamics of the capacitor and simply represents the battery as a voltage source behind a resistor, as in \cite{bazargan2014stability}.

In the per unit case, the DC-side base power is the same the AC-side. The DC-side base voltage, however, is two times the AC-side peak line-to-neutral base voltage. This is done to obtain an AC-side voltage of $1.0$ p.u. from the a DC-side voltage of $1.0$ p.u. at unity modulation ratio \cite{yazdani2010voltage}. The saturation of the PWM modulation index is implemented similar to \cite{pico2019transient} as
\begin{equation}\label{eq:sat}
    \boldsymbol{v}_{m} = \dfrac{\min\{\vert \vert \bar{\boldsymbol{v}}_{m}\vert \vert_{2}, v_{\dc}\}}{\vert \vert \bar{\boldsymbol{v}}_{m}\vert \vert_{2}}\bar{\boldsymbol{v}}_{m},
\end{equation}
where $\bar{\boldsymbol{v}}_{m}$ is given by \eqref{eq:srf_i} and $\vert \vert \bar{\boldsymbol{v}}_{m}\vert \vert_{2}$ is
\begin{equation} \label{eq:output_v_mag}
    \vert \vert \bar{\boldsymbol{v}}_{m}\vert \vert_{2} = \sqrt{\bar{v^d_m}^2 +\bar{v^q_m}^2}.
\end{equation}


%% file: method.tex
\section{Methodology}
\label{sec:meth}

In this section we outline a methodology for choosing the control gains of the DC/DC converter, in order to understand and improve the dynamical behavior of the DC-side of the CIG. To this end, we use a linearized model of our system, as presented in Section~\ref{small_signal} and identify a set of gains that result in stable operating points. Subsequently, in Section~\ref{large_signal}, we determine the gains from this set which optimize the dynamical performance of the DC/DC controller under large disturbances. To account for the discrete nature of the DC/DC controller we utilize a Pade approximation of the associated dead-time delay. The average output performance for a step input of a $2^\text{\it nd}$ and $3^\text{\it rd}$-order approximation is used to model the dead-time of the DC/DC controller.  

\subsection{Small-signal tuning}\label{small_signal}
We express the non-linear differential equations \eqref{eq:outer_loop}-\eqref{eq:dc_elec_eqs} as
\begin{equation}
    \dot{\bm{x}} = \bm{f}(\bm{x},\bm{u},\bm{w}),
\end{equation}
where $\bm{x},\bm{u},\bm{w}$ correspond to the states, inputs, and external disturbances (loads), respectively. For the purpose of analysis, we linearize this system around an equilibrium point $(\bm{x}_{eq},\bm{u}_{eq},\bm{w}_{eq})$ to obtain a resultant linear system 
\begin{equation}
    \Delta \dot{x} = A\Delta x + B \Delta w,
\end{equation}
where the matrices $A$ and $B$ are evaluated as
\begin{equation}
    A = \left.\dfrac{\partial \bm{f}}{\partial \bm{x}}\right\vert_{(\bm{x}_{eq},\bm{u}_{eq},\bm{w}_{eq})}, \quad  B = \left.\dfrac{\partial \bm{f}}{\partial \bm{w}}\right\vert_{(\bm{x}_{eq},\bm{u}_{eq},\bm{w}_{eq})}.
\end{equation}

The task of small-signal tuning involves finding a set of DC-side control gains
\begin{equation}
K^{\!_\textup{\it DC}} = [K_{p}^{v_{\!_\textup{\it DC}}},K_{i}^{v_{\!_\textup{\it DC}}},K_{p}^{i_{\!_\textup{\it DC}}},K_{i}^{i_{\!_\textup{\it DC}}}, K_\textup{\it pred}] 
\end{equation}
which satisfy some pre-specified design requirements, e.g.,
\begin{subequations}\label{eq:design_spec}
\begin{align}
   \Re[\lambda_{i}(A(K^{\!_\textup{\it DC}}))] \leq \lambda_\textup{\it crit} \quad \forall i, \\
   \zeta_{i} \geq \zeta_\textup{\it crit} \quad \forall i, \\
   K^{\!_\textup{\it DC}}_\textup{\it min} \leq K^{\!_\textup{\it DC}} \leq K^{\!_\textup{\it DC}}_\textup{\it max},
\end{align}
\end{subequations}
where $\lambda$ and $\zeta$ correspond to the eigenvalues and the damping ratio of the linearized model respectively, $\lambda_{crit}$ and $\zeta_{crit}$ are design requirements, and $K^{\!_\textup{\it DC}}_\textup{\it max}$ and $K^{\!_\textup{\it DC}}_\textup{\it min}$ represent some pre-specified limits on the control gains. We denote this set of all permissible gains by the set $\Gamma$.

\subsection{Large-signal tuning}\label{large_signal}
On identifying a set of suitable small-signal gains $\Gamma$, an exhaustive search over this set is performed to optimize the dynamical performance of the full non-linear system when it is subject to large disturbances, e.g., large load step changes. In particular, we seek to identify the set of gains that minimize the DC voltage deviation from its set point. This can be expressed mathematically as minimizing the $\ell_{2}$-norm
\vspace{0.1em}
\begin{equation}\label{eq:opt}
\begin{split}
    \min_{K^{\!_\textup{\it DC}} \in \Gamma} \,\,&\vert \vert v_{\!_\textup{\it DC}}^{*}-v_{\!_\textup{\it DC}}(t)\vert \vert _{2}^{2}\\
    \text{subject to} \,\,&\eqref{eq:outer_loop}-\eqref{eq:dc_elec_eqs}\\
    &p_{l}(t_{0})=p_{l}, \, p_{l}(t)=p_{l}+\Delta p_{l},
\end{split}
\end{equation}
\vspace{0.1em}
where $p_{l}$ is the nominal active power load and $\Delta p_{l}$ represents a disturbance in the form of a step-change increase in the load. We first optimize the DC/DC control gains with $K_\textup{\it pred}=0$ and then benchmark the improved dynamical performance for cases where $K_\textup{\it pred}\neq0$. Section ~\ref{sec:results} discusses the design requirements and disturbance used in \eqref{eq:design_spec} and \eqref{eq:opt} respectively.

%% file: results.tex
\section{Results} \label{sec:results}

The simulations  are  performed  using  the  Julia programming language.  The ModelingToolkit.jl package is used to construct the non-linear system and perform the Jacobian evaluations. The power rating of the VSC is $200$ kVA and the parameters are taken from  \cite{DArco2014} while parameters for the DC-side are presented in Appendix \ref{sec:app}. The controller design parameters used for both the small-signal and large-signal tuning are shown in Table~\ref{tab:tuning_parameters}. The small-signal parameter search is carried out by a grid search with step size 0.5.  All the analysis presented here is available on Github\footnote{https://github.com/Energy-MAC/DCSideBatteryModeling}. 

\begin{table}[htp]
\normalsize
\renewcommand{\arraystretch}{1.3}
\caption{\label{tab:tuning_parameters}Controller tuning parameters}
\centering
\begin{tabular}{c||c|c|c|c|c}
 Specification& $\lambda_\textup{\it crit}$ & $\zeta_\textup{\it crit}$ & $K^{\!_\textup{\it DC}}_\textup{\it max}$ &$K^{\!_\textup{\it DC}}_\textup{\it min}$ & $\Delta p_{l}$\\
 \hline
Value & $-3$ & $0.35$ & $10$ & $0$ & $0.5$ p.u.
\end{tabular}
\end{table}

\subsection{Comparing BESS Models}
In Fig.~\ref{fig:untuned_gains}, we compare the DC-side voltage of the three BESS models, i.e., $4^\textup{\it th}$, $2^\textup{\it nd}$, and $0^\textup{\it th}$-orders for non-optimized controller gains under a load step change of $\Delta p_{l}= 0.5$\,p.u. We observe that all models are in agreement regarding the dynamical response (also true for different controller gains). Further, we note that the results here only apply to a Lithium-ion based BESS for the parameters from \cite{hart2014modeling,rahmoun2016mathematical}. For the case of compressed air storage with associated mechanical dynamics and redox flow batteries, with different underlying chemistry; a higher order model representation may be necessary.   

\input{figs/untuned_gains_model_comparsion}

\subsection{Impact of one-step predictor}
In order to examine the improvement in controller performance by inclusion of the AC-side measurements, we examine the response of the system to a load step change of $\Delta p_{l}=0.5$ p.u. for varying values of $K_\textup{\it pred}$. Fig.~\ref{fig:kpred_plots} shows the DC voltage for three different values of $K_\textup{\it pred}$. We observe up to a $\sim 10$\% reduction in the maximum DC voltage error after including the AC measurements. This reduction, achieved using existing measurements readily available in the VSC control loop, offers a means to reduce the severity of transients across the DC-link capacitor and reduce overloading in the event of over-voltage, two of the dominant reasons for premature failure \cite{yang2011industry}. 
\input{figs/increasing_kpred}
\input{figs/heatmap}

Fig.~\ref{fig:heat_map} further explores the performance of the optimized controller for varying DC-link capacitor sizing. We see that the inclusion of AC-side measurements does offer some improvement, however, due to the saturation behavior of the DC/DC boost converter this improvement is upper-bounded. Therefore, while the AC-side measurement improves the dynamical performance and reduces transient behavior across the capacitor, it only offers a modest reduction in DC-link capacitor sizing for a pre-specified $\ell_{2}$ norm performance requirement.  

In order to understand the limiting factor in the response of the BESS to regulate the DC voltage, we examine the battery current $i_{b}$, shown in Fig. \ref{fig:batt_current}. We can see that the dead-time of the DC/DC controller only accounts for a small proportion of the delay in the response. The majority of the delay is due to the dynamics of the DC/DC inductor, in this case $3$ mH. While this is a physical design limitation and there exist approaches to minimize the required inductance to improve dynamic response, e.g., increasing the switching frequency \cite{hart2014modeling} or operating in discontinuous conduction mode \cite{bazargan2014stability}, these design questions are beyond the scope of this work.
\input{figs/batt_current}

\subsection{Examining VSC behavior}
One additional benefit of including the AC-side measurements, and consequently, better regulation of the DC voltage, is the opportunity to reduce the DC-link capacitor size without saturating the PWM converter. 

For the simulations considered in this paper with grid-forming inverter control gains from \cite{DArco2014}, the saturation of the PWM converter was avoided in all cases examined. Fig.~\ref{fig:saturation} shows both the DC voltage $V{\dc}$ and magnitude of the modulated AC-side voltage $\vert \vert \bm{v}_{m} \vert \vert $, for the case of  $K_\textup{\it pred}=2$. The inner control loops of the grid-forming VSC respond on a faster timescale to reduce the magnitude of the AC modulated voltage and thereby, significantly reduce the risk of saturating the modulation index of the VSC. The outer control loops of the VSC then re-adjust the set points to restore the voltage to an acceptable operating level. While saturation was not an issue in this set up, it may be an issue for different parameterizations and/or disturbances. 

\input{figs/saturation}

%% file: figs/untuned_gains_model_comparsion.tex
\begin{figure}
\begin{tikzpicture}
\begin{axis}[
width=3.5in,
height=2.2in,
label style={font=\footnotesize},
tick label style={font=\footnotesize},
ytick={0.9,0.95,1,1.05,1.1, 1.15},
xticklabel={\pgfmathparse{int(round(\tick*10^3))}$\pgfmathprintnumber{\pgfmathresult}$},
        colormap name=viridis,
		xlabel= Time ($10^{-3}$s),
		ylabel=$V_{{\!_\textup{\it DC}}}$ (p.u.),
        xmajorgrids=true,
        ymajorgrids=true,
        grid style=dashed,
		xmin=0,
        xmax=0.05,
        ymin=0.88,
        ymax=1.17,
        scaled x ticks = false, 
        legend style={at={(0.95,0.05)},anchor=south east,draw=none, font=\footnotesize}]
\addplot [color=mycolor1, line width=2pt] table [x index = {0}, y index = {13}, col sep=comma] {data/untuned_4th_2mF.csv};
\addplot [color=mycolor2, line width=2pt, dashed] table [x index = {0}, y index = {13}, col sep=comma] {data/untuned_2nd_2mF.csv};
\addplot [color=mycolor3, line width=2pt, dotted] table [x index = {0}, y index = {13}, col sep=comma] {data/untuned_0th_2mF.csv};
\legend{$4^\textup{\it th}-$order BESS model, $2^\textup{\it nd}-$order BESS model, $0^\textup{\it th}-$order BESS model,}
\end{axis}
\end{tikzpicture}
\caption{BESS response comparison for non-optimized gains.} \label{fig:untuned_gains}
\end{figure}
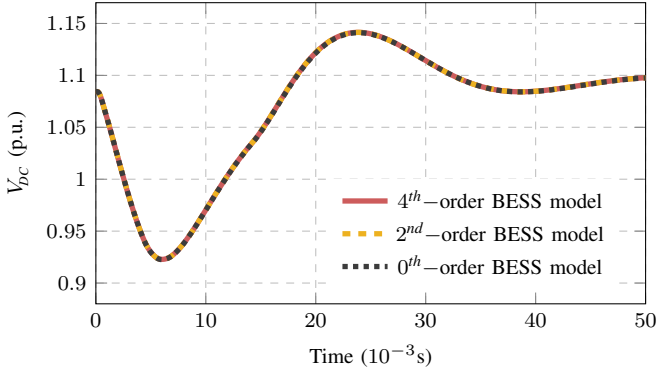

%% file: figs/increasing_kpred.tex
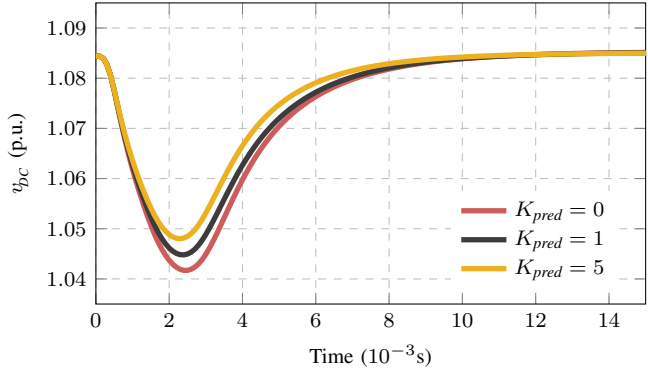
\begin{figure}
\begin{tikzpicture}
\begin{axis}[
width=3.5in,
height=2.2in,
label style={font=\footnotesize},
legend cell align=left,
tick label style={font=\footnotesize},
ytick={1.04,1.05,1.06,1.07,1.08, 1.09},
xticklabel={\pgfmathparse{int(round(\tick*10^3))}$\pgfmathprintnumber{\pgfmathresult}$},
        colormap name=viridis,
		xlabel= Time ($10^{-3}$s),
		ylabel=$v_{\dc}$ (p.u.),
        xmajorgrids=true,
        ymajorgrids=true,
        grid style=dashed,
        ymin=1.035,
        ymax=1.095,
		xmin=0,
        xmax=0.015,
        scaled x ticks = false, 
        legend style={at={(0.95,0.05)},anchor=south east,draw=none, font=\footnotesize}]
\addplot [color=mycolor1, line width=2pt] table [x index = {0}, y index = {13}, col sep=comma] {data/kpred_0.csv};
\addplot [color=mycolor3, line width=2pt] table [x index = {0}, y index = {13}, col sep=comma] {data/kpred_1.csv};
\addplot [color=mycolor2, line width=2pt] table [x index = {0}, y index = {13}, col sep=comma] {data/kpred_5.csv};
\legend{$K_\textup{\it pred}=0$, $K_\textup{\it pred}=1$, $K_\textup{\it pred}=5$}
\end{axis}
\end{tikzpicture}
\caption{Optimized controller performance with one-step predictor.} \label{fig:kpred_plots}
\end{figure}

%% file: figs/heatmap.tex
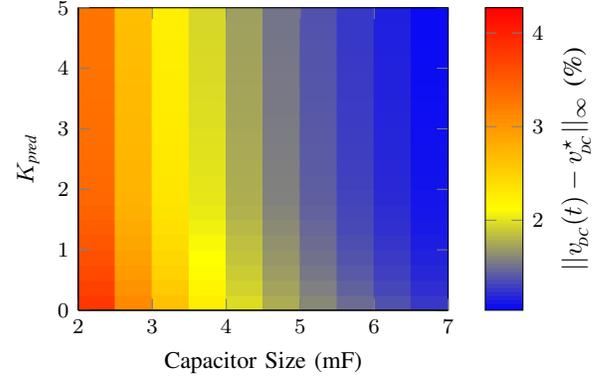
\begin{figure}[!h]
\centering
\begin{tikzpicture} 
    \begin{axis}[
    xlabel=Capacitor Size (mF),
    ylabel=$K_\textup{\it pred}$,
    small,view={0}{90},
    colorbar,
    colorbar style={ylabel=$\vert\vert v_{ \dc}(t)-v_{\dc}^{\star}\vert\vert_{\infty}$ (\%)}
]
       \addplot3 [surf,
       shader=flat,
       patch type=bilinear,
        ]
        table {data/heat_map_data.dat};
    \end{axis}
\end{tikzpicture}
\caption{Comparing maximum $v_{\dc}$ deviation for varying DC-link capacitor sizing.} \label{fig:heat_map}
\end{figure}

%% file: figs/batt_current.tex
\begin{figure}
\begin{tikzpicture}
\begin{axis}[
width=3.5in,
height=2.4in,
label style={font=\footnotesize},
tick label style={font=\footnotesize},
xticklabel={\pgfmathparse{int(round(\tick*10^3))}$\pgfmathprintnumber{\pgfmathresult}$},
        colormap name=viridis,
		xlabel= Time ($10^{-3}$s),
		ylabel=$i_{b}$ (p.u.),
        xmajorgrids=true,
        ymajorgrids=true,
        grid style=dashed,
		xmin=0,
        xmax=0.005,
        ymin=0.3,
        ymax=1.65,
        scaled x ticks = false, 
        legend style={at={(0.95,0.05)},anchor=south east,draw=none, font=\footnotesize}]
\addplot [color=mycolor1, line width=2pt] table [x index = {0}, y index = {14}, col sep=comma] {data/kpred_0.csv};
\addlegendentry{$K_\textup{\it pred}=0$}; 
\addplot [color=mycolor2, line width=2pt] table [x index = {0}, y index = {14}, col sep=comma] {data/kpred_5.csv};
\addlegendentry{$K_\textup{\it pred}=5$}; 
\addplot[fill=gray, fill opacity = 0.2, draw = none,area legend] coordinates {(0,2) (0,0.2) (0.0003125,0.1) (0.0003125,2)};
\addlegendentry[align=left]{controller delay}; 
\end{axis}
\end{tikzpicture}
\caption{Battery current profile with one-step predictor.} \label{fig:batt_current}
\end{figure}
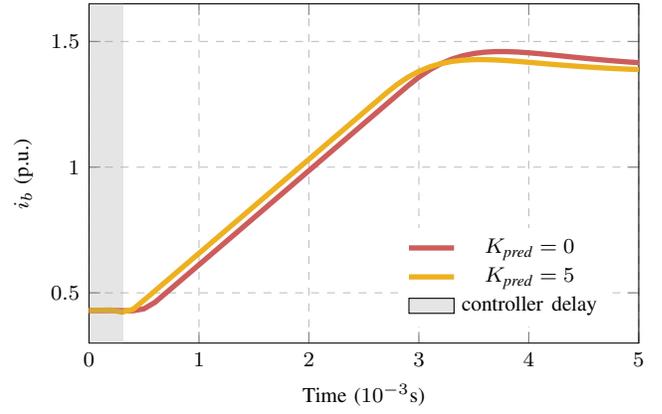

%% file: figs/saturation.tex

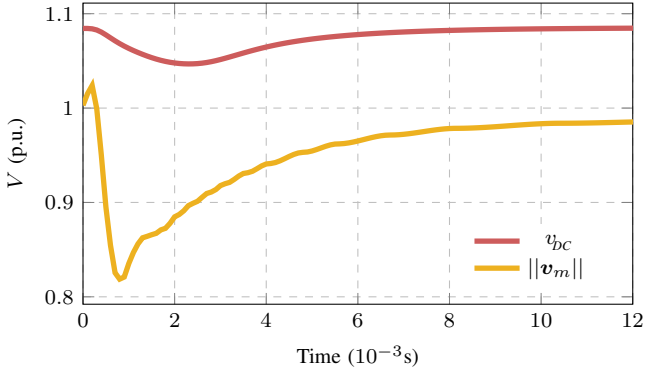
\begin{figure}
\begin{tikzpicture}
\begin{axis}[
width=3.5in,
height=2.2in,
label style={font=\footnotesize},
tick label style={font=\footnotesize},
xticklabel={\pgfmathparse{int(round(\tick*10^3))}$\pgfmathprintnumber{\pgfmathresult}$},
        colormap name=viridis,
		xlabel= Time ($10^{-3}$s),
		ylabel=$V$ (p.u.),
        xmajorgrids=true,
        ymajorgrids=true,
        grid style=dashed,
		xmin=0,
        xmax=0.012,
        scaled x ticks = false, 
        legend style={at={(0.95,0.05)},anchor=south east,draw=none, font=\footnotesize}]
\addplot [color=mycolor1, line width=2pt] table [x index = {0}, y index = {13}, col sep=comma] {data/kpred_2.csv};
\addplot [color=mycolor2, line width=2pt] table [x index = {0}, y index = {24}, col sep=comma] {data/kpred_2.csv};
\legend{$v_{\dc}$, $\vert\vert \bm{v}_{m} \vert \vert\ $}
\end{axis}
\end{tikzpicture}
\caption{DC voltage and AC modulated voltage for $K_\textup{\it pred}=2$.} \label{fig:saturation}
\end{figure}

%% file: conclusions.tex
\section{Conclusion} \label{sec:conc}

This work focused on modeling and control of a BESS DC source grid forming VSC. On the modeling side the DC/DC inductor was observed to be the dominant component dictating the dynamical behavior. A 4$^\textup{\it th}$, $2^\textup{\it nd}$, and $0^\textup{\it th}$-order model of a BESS was examined and it was found that all three models were in agreement for the considered disturbances. For the DC/DC controller, it was found that the inclusion of readily available AC-side measurements into the DC/DC converter control loop could reduce DC voltage deviations by up to $\sim10$\% during large step changes, thereby potentially reducing the risk of premature failure of the DC-link capacitor. Future work will focus on the behavior of these controllers under asymmetrical gird faults, additional DC-source technologies as well as further consideration of how fast inner-control loops of the VSC which may help alleviate the potential for saturation of the VSC modulation index. 

%% file: appendix.tex
\appendix \label{sec:app}

Table \ref{tab:dc_side} lists the parameter values used for the DC-side model for simulations \cite{hart2014modeling, rahmoun2016mathematical}

\begin{table}[htp]
\normalsize
\renewcommand{\arraystretch}{1.25}
\caption{\label{tab:dc_side}DC-side parameters}
\centering
 \begin{threeparttable}
\begin{tabular}{c|c|c|c|c|c}
$f^{s}_\textup{\it DC/DC}$& $c_{\dc}$& $l_{\dc}$& $r_{b0}$ & $r_{b1}$ & $r_{b2}$\\
\hline
$3.2$\,kHz & $2$\,mF & $3$\,mH & $1.5$\,m$\Omega$ & $95$\,m$\Omega$ & $0.4$\,m$\Omega$
\end{tabular}
 \end{threeparttable}
\vspace{0.05in}

\begin{tabular}{c|c|c|c|c|c}
$r_{b3}$ & $r_{b4}$ & $l_{b1}$ & $l_{b2}$ & $c_{b1}$ & $c_{b2}$ \\
\hline
$2.2$\,m$\Omega$ & $0.55$\,m$\Omega$ & $35$\,nH & $15$\,nH & $0.55$\,F & $22.7$\,kF
\end{tabular}
\end{table}

%% file: main.bbl
\begin{thebibliography}{10}
\providecommand{\url}[1]{#1}
\csname url@samestyle\endcsname
\providecommand{\newblock}{\relax}
\providecommand{\bibinfo}[2]{#2}
\providecommand{\BIBentrySTDinterwordspacing}{\spaceskip=0pt\relax}
\providecommand{\BIBentryALTinterwordstretchfactor}{4}
\providecommand{\BIBentryALTinterwordspacing}{\spaceskip=\fontdimen2\font plus
\BIBentryALTinterwordstretchfactor\fontdimen3\font minus
  \fontdimen4\font\relax}
\providecommand{\BIBforeignlanguage}[2]{{%
\expandafter\ifx\csname l@#1\endcsname\relax
\typeout{** WARNING: IEEEtran.bst: No hyphenation pattern has been}%
\typeout{** loaded for the language `#1'. Using the pattern for}%
\typeout{** the default language instead.}%
\else
\language=\csname l@#1\endcsname
\fi
#2}}
\providecommand{\BIBdecl}{\relax}
\BIBdecl

\bibitem{chen2015resilient}
C.~Chen, J.~Wang, F.~Qiu, and D.~Zhao, ``Resilient distribution system by
  microgrids formation after natural disasters,'' \emph{IEEE Transactions on
  smart grid}, vol.~7, no.~2, pp. 958--966, 2015.

\bibitem{li2017networked}
Z.~Li, M.~Shahidehpour, F.~Aminifar, A.~Alabdulwahab, and Y.~Al-Turki,
  ``Networked microgrids for enhancing the power system resilience,''
  \emph{Proceedings of the IEEE}, vol. 105, no.~7, pp. 1289--1310, 2017.

\bibitem{d2015small}
S.~D’Arco, J.~A. Suul, and O.~B. Fosso, ``Small-signal modeling and
  parametric sensitivity of a virtual synchronous machine in islanded
  operation,'' \emph{International Journal of Electrical Power \& Energy
  Systems}, vol.~72, pp. 3--15, 2015.

\bibitem{UrosStability}
\BIBentryALTinterwordspacing
U.~Markovic, O.~Stanojev, E.~Vrettos, P.~Aristidou, D.~Callaway, and G.~Hug,
  ``{Understanding Stability of Low-Inertia Systems},'' (in preparation).
  [Online]. Available: \url{engrxiv.org/jwzrq}
\BIBentrySTDinterwordspacing

\bibitem{bazargan2014stability}
D.~Bazargan, S.~Filizadeh, and A.~M. Gole, ``Stability analysis of
  converter-connected battery energy storage systems in the grid,'' \emph{IEEE
  Transactions on Sustainable Energy}, vol.~5, no.~4, pp. 1204--1212, 2014.

\bibitem{bazargan2018reduced}
D.~Bazargan, B.~Bahrani, and S.~Filizadeh, ``Reduced capacitance battery
  storage dc-link voltage regulation and dynamic improvement using a
  feedforward control strategy,'' \emph{IEEE Transactions on Energy
  Conversion}, vol.~33, no.~4, pp. 1659--1668, 2018.

\bibitem{wang2013toward}
H.~Wang, M.~Liserre, and F.~Blaabjerg, ``Toward reliable power electronics:
  Challenges, design tools, and opportunities,'' \emph{IEEE Industrial
  Electronics Magazine}, vol.~7, no.~2, pp. 17--26, 2013.

\bibitem{yang2011industry}
S.~Yang, A.~Bryant, P.~Mawby, D.~Xiang, L.~Ran, and P.~Tavner, ``An
  industry-based survey of reliability in power electronic converters,''
  \emph{IEEE transactions on Industry Applications}, vol.~47, no.~3, pp.
  1441--1451, 2011.

\bibitem{lee2013dc}
W.-J. Lee and S.-K. Sul, ``Dc-link voltage stabilization for reduced dc-link
  capacitor inverter,'' \emph{IEEE Transactions on industry applications},
  vol.~50, no.~1, pp. 404--414, 2013.

\bibitem{elkhatib2018evaluation}
M.~E. Elkhatib, W.~Du, and R.~H. Lasseter, ``Evaluation of inverter-based grid
  frequency support using frequency-watt and grid-forming pv inverters,'' in
  \emph{2018 IEEE Power \& Energy Society General Meeting (PESGM)}.\hskip 1em
  plus 0.5em minus 0.4em\relax IEEE, 2018, pp. 1--5.

\bibitem{ferraz2018high}
P.~K.~P. Ferraz, R.~Schmidt, D.~Kober, and J.~Kowal, ``A high frequency model
  for predicting the behavior of lithium-ion batteries connected to fast
  switching power electronics,'' \emph{Journal of Energy Storage}, vol.~18, pp.
  40--49, 2018.

\bibitem{DArco2014}
S.~D'Arco, J.~A. Suul, and O.~B. Fosso, ``Small-signal modelling and parametric
  sensitivity of a virtual synchronous machine,'' in \emph{PSCC}, 2014.

\bibitem{UrosISGTeurope}
U.~Markovic, J.~Vorwerk, P.~Aristidou, and G.~Hug, ``Stability analysis of
  converter control modes in low-inertia power systems,'' in \emph{IEEE
  Innovative Smart Grid Technologies - Europe (ISGT-Europe)}, Oct 2018.

\bibitem{UrosEnergyCon}
R.~Ofir, U.~Markovic, P.~Aristidou, and G.Hug, ``Droop vs. virtual inertia:
  Comparison from the perspective of converter operation mode,'' in \emph{IEEE
  International Energy Conference (ENERGYCON)}, June 2018.

\bibitem{gu2006dc}
B.-G. Gu and K.~Nam, ``A dc-link capacitor minimization method through direct
  capacitor current control,'' \emph{IEEE Transactions on Industry
  Applications}, vol.~42, no.~2, pp. 573--581, 2006.

\bibitem{zhao2006dc}
L.~Zhao and J.~Qian, ``Dc-dc power conversions and system design considerations
  for battery operated system,'' \emph{Texas Instruments}, 2006.

\bibitem{uno2011influence}
M.~Uno and K.~Tanaka, ``Influence of high-frequency charge--discharge cycling
  induced by cell voltage equalizers on the life performance of lithium-ion
  cells,'' \emph{IEEE Transactions on vehicular technology}, vol.~60, no.~4,
  pp. 1505--1515, 2011.

\bibitem{rahmoun2016mathematical}
A.~Rahmoun, A.~Armstorfer, J.~Helguero, H.~Biechl, and A.~Rosin, ``Mathematical
  modeling and dynamic behavior of a lithium-ion battery system for microgrid
  application,'' in \emph{2016 IEEE International Energy Conference
  (ENERGYCON)}.\hskip 1em plus 0.5em minus 0.4em\relax IEEE, 2016, pp. 1--6.

\bibitem{waag2013experimental}
W.~Waag, S.~K{\"a}bitz, and D.~U. Sauer, ``Experimental investigation of the
  lithium-ion battery impedance characteristic at various conditions and aging
  states and its influence on the application,'' \emph{Applied energy}, vol.
  102, pp. 885--897, 2013.

\bibitem{hart2014modeling}
P.~Hart, P.~Kollmeyer, L.~Juang, R.~Lasseter, and T.~Jahns, ``Modeling of
  second-life batteries for use in a certs microgrid,'' in \emph{2014 Power and
  Energy Conference at Illinois (PECI)}.\hskip 1em plus 0.5em minus 0.4em\relax
  IEEE, 2014, pp. 1--8.

\bibitem{wang2014nonlinear}
C.~Wang, X.~Li, L.~Guo, and Y.~W. Li, ``A nonlinear-disturbance-observer-based
  dc-bus voltage control for a hybrid ac/dc microgrid,'' \emph{IEEE
  Transactions on Power Electronics}, vol.~29, no.~11, pp. 6162--6177, 2014.

\bibitem{yazdani2010voltage}
A.~Yazdani and R.~Iravani, \emph{Voltage-sourced converters in power systems:
  modeling, control, and applications}.\hskip 1em plus 0.5em minus 0.4em\relax
  John Wiley \& Sons, 2010.

\bibitem{pico2019transient}
H.~N.~V. Pico and B.~B. Johnson, ``Transient stability assessment of
  multi-machine multi-converter power systems,'' \emph{IEEE Transactions on
  Power Systems}, vol.~34, no.~5, pp. 3504--3514, 2019.

\end{thebibliography}
